\documentclass[english]{paper}
\usepackage[T1]{fontenc}
\usepackage[latin9]{inputenc}
\usepackage{amsmath}
\usepackage{amssymb}
\usepackage{graphicx}
\usepackage[authoryear]{natbib}

\makeatletter
\newcommand{\lyxaddress}[1]{
\par {\raggedright #1
\vspace{1.4em}
\noindent\par}
}

\makeatother

\usepackage{babel}
\begin{document}

\title{Response Selection Using Neural Phase Oscillators}

\author{J. Acacio de Barros$^{*}$, and G. Oas$^{\dagger}$}

\maketitle

\lyxaddress{$^{*}$Liberal Studies, San Francisco State University, San Francisco,
CA, 94132\\
$^{\dagger}$CSLI, Stanford University, Stanford, CA 94305}
\begin{abstract}
In a recent paper, Suppes, de Barros, \& Oas \citeyearpar{suppes_phase-oscillator_2012}
used neural oscillators to create a model, based on reasonable neurophysiological
assumptions, of the behavioral stimulus-response (SR) theory. In this
paper, we describe the main characteristics of the model, emphasizing
its physical and intuitive aspects. 
\end{abstract}

\section{Introduction}

It is an honor for Acacio de Barros and Gary Oas to participate in
a festschrift for Pat Suppes. It is especially rewarding to do so
with a paper where we discuss our most recent work with Pat, a model
of brain processes using neural oscillators. We are, as Pat would
say, ``true blue physicists.'' So, for us, collaborating with Pat
on this truly interdisciplinary paper is an example not only of his
intellectual influence, but also of his friendship and mentorship.
We are happy to dedicate this paper to Pat. Happy Birthday Pat! 

The work we present here started more than ten years ago, when JAB
and Pat begun thinking about how to model in a physically plausible
way collections of neurons in terms of oscillators. In one of his
known intuitions, Pat kept insisting that the brain ``gotta use oscillators.''
Of course, as is often the case, his ``intuition'' was based on
hard work and detailed empirical data that he collected working on
the EEG of words and sentences. Nevertheless, as we kept trying to
make our model work (and we had many failures, and some successes;
see \citet{vassilieva_learning_2011} for an example), Pat kept insisting:
we should understand the brain computations with oscillators. I am
pleased to say that, despite my initial skepticism, we now have a
model that we feel is not only grounded on neurophysiologically sound
evidence, but that also reproduces quite well some empirical behavioral
data (Suppes, de Barros, \& Oas, 2012). 

In this paper we attempt to describe the main features this model
by focusing on the physical processes underlying the neural computations.
We chose to do so for the following reasons. First, because of its
interdisciplinarity, our model requires concepts from many different
areas (neurophysiology, physics, psychology, etc). Such concepts are
not complex, but are often unfamiliar to most researchers. Second,
we are confident that our model is relevant to cognitive psychologists,
as it may explain some mathematical models showing good empirical
fit \citep{de_barros_joint_2012,de_barros_quantum-like_2012}. So,
we believe that this paper can provide a clearer and intuitive view
of the main physical features of our model for those thinking about
applying it, supplementing the discussions found in Suppes, de Barros,
\& Oas \citeyearpar{suppes_phase-oscillator_2012}.

Let us start our discussion with the broad problem of understanding
how the brain processes information. This is perhaps the most challenging
current scientific endeavors, mainly due to the fact that our brain
is tremendously complicated, as it is constituted of  many different
components that are, by themselves, complex, but that also seem to
sometimes interact holistically with each other. Among the approaches
to try and understand the brain, the most prominent ones are the top-down
and bottom-up. In the top-down approach, we start with the higher-level
functions and go to their underlying mechanisms. An example of such
approach would be the field of cognitive neuroscience, where often
one starts with experiments in cognitive psychology and tries to understand
them from principles in neuroscience \citep{adolphs_cognitive_2003}.
In the bottom-up approach, one tries to start with neurophysiology,
and by studying how each elementary component works, one tries to
see how higher functions arise from such components or their interaction
\citep{kandel_principles_2000}. 

Each of those approaches have their shortcomings. For example, one
of the main issues is what we may call a problem of scale. When trying
to understand a complex system, the first question that arises is
how detailed we need to be. In the case of the brain, some researchers
say that we need to go all the way down to the chemical reactions
in the synapses. Others argue that individual neurons hold the key
to understanding brain computation. Yet another view is that collections
of neurons are important. So, when trying to understand how the brain
works, our first problem is where to begin. Regardless of what scale
is chosen and where we start, ultimately we would need to understand
the whole process if we were to claim to have understood the brain. 

The main problem with connecting a higher scale with a lower one is
due to its complexity. For example, evidence exists that higher cognitive
processes involve tens to hundreds of thousands of neurons, interacting
with each other in very complex ways. Modeling such processes require
the use of powerful computers. But, even when a model is shown to
work from the underlying neuronal dynamics, the use of massive computer
simulations helps little in understanding, in an intuitive or conceptual
way, what is actually happening. The system is simply too complex. 

To deal with the issue of complexity, different approaches can be
taken. One possible route is to find physically plausible arguments
that impose constraints on the system's dynamics, therefore reducing
it to fewer degrees of freedom. This is the approach taken by Suppes,
de Barros, \& Oas (2012). In their paper, a large number of independent
neurons was modeled by a single dynamical parameter determined by
the phase of a neural oscillator. They then showed that under certain
reasonable assumptions, the main characteristics of behavioral stimulus-response
(SR) theory could be described by neural oscillators. The use of neural
oscillators thus provided a significant reduction on the number of
degrees of freedom, allowing for the physical interpretation of many
different parameters in the model. 

In this paper we present the work of Suppes, de Barros, \& Oas (2012),
with emphasis on the physics and intuition behind the model. Our goal
is to make this model more understandable, as many of the concepts
used in our previous paper are not well-known to certain audiences.
For example, while all physicists have an excellent knowledge of oscillations
and interference and could easily follow the arguments leading from
neurons to oscillators, only a few would feel comfortable with the
mathematical learning theories used. Neuroscientists, on the other
hand, would probably feel at home with neurons and learning theories,
but not so much with oscillators and interference. Neither would most
psychologists. Here we focus on the intuitions behind the physics,
with the hopes that, in conjunction with the oscillator model, psychologists
and neuroscientists could benefit more from the insights gained.

\section{A Brief Review of SR theory}

Stimulus-response theory (or SR theory; see \citet{suppes_markov_1960})
is one of the most successful behavioral learning theories in psychology.
Though it has decreased in importance in current psychology, we chose
to model SR theory for the following reasons. First, it is based on
a rigid trial structure, which permits its concepts to be formally
axiomatized, resulting in many important non-trivial but illuminating
representation theorems \citep{suppes_representation_2002}. In fact,
the theory is rich enough to represent language in it. Second, despite
its few parameters (the learning probability $c$ and the number of
stimuli), it has been shown to fit well to empirical data in a variety
of experiments. Finally, as we showed in Suppes, de Barros, \& Oas
(2012), SR theory seems to have natural counterparts at a neuronal
level, and is, in some sense still used by neuroscientists (though,
sadly, not in its mathematical form). 

Here we present the mathematical version of SR theory for a continuum
of responses, formalized in terms of a stochastic process (we follow
Suppes, de Barros, \& Oas, 2012). Let $\left(\Omega,{\cal F},P\right)$
be a probability space, and let $\mathbf{Z}$, $\mathbf{S}$, $\mathbf{R}$,
and $\mathbf{E}$ be random variables, with $\mathbf{Z}:\Omega\rightarrow E^{\left|S\right|}$
$\mathbf{S}:\Omega\rightarrow S$, $\mathbf{R}:\Omega\rightarrow R$,
and $\mathbf{E}:\Omega\rightarrow E$, where $S$ is the set of stimuli,
$R$ the set of responses, and $E$ the set of reinforcements. Then
a trial in SR theory has the following structure: 
\begin{equation}
\mathbf{Z}_{n}\rightarrow\mathbf{S}_{n}\rightarrow\mathbf{R}_{n}\rightarrow\mathbf{E}_{n}\rightarrow\mathbf{Z}_{n+1}.\label{eq:SR-trial}
\end{equation}
The trial structure works the following way. Trial $n$ starts with
a certain state of conditioning and a sampled stimulus. Once a stimulus
is sampled, a response is computed according to the state of conditioning.
Then, reinforcement follows, which can lead (with probability $c$)
to a new state of conditioning for trial $n+1$ . In more detail,
at the beginning of a trial, the state of conditioning is represented
by the random variable $\mathbf{Z}_{n}=\left(z_{1}^{\left(n\right)},\ldots,z_{m}^{\left(n\right)}\right)$.
The vector $\left(z_{1}^{\left(n\right)},\ldots,z_{m}^{\left(n\right)}\right)$
associates to each stimuli $s_{i}\in S$, $i=1,\ldots,m$, where $m=\left|S\right|$
is the cardinality of $S$, a value $z_{i}^{\left(n\right)}$ on trial
$n$. Once a stimulus $\mathbf{S}_{n}=s_{i}$ is sampled with probability
$P\left(\mathbf{S}_{n}=s_{i}|s_{i}\epsilon S\right)=\dfrac{1}{m}$,
its corresponding $z_{i}^{\left(n\right)}$ determines the probability
of responses in $R$ by the probability distribution $K\left(r|z_{i}^{\left(n\right)}\right)$,
i.e. $P\left(a_{1}\leq\mathbf{R}_{n}\leq a_{2}|\mathbf{S}_{n}=s_{i},\mathbf{Z}_{n,i}=z_{i}^{\left(n\right)}\right)=\int_{a_{1}}^{a_{2}}k\left(x|z_{i}^{\left(n\right)}\right)dx$,
where $k\left(x|z_{i}^{\left(n\right)}\right)$ is the probability
density associated to the distribution, and where $\mathbf{Z}_{n,i}$
is the $i$-th component of the vector $\left(z_{1}^{\left(n\right)},\ldots,z_{m}^{\left(n\right)}\right)$.
The probability distribution $K\left(r|z_{i}^{\left(n\right)}\right)$
is the smearing distribution, and it is determined by its variance
and mode $z_{i}^{\left(n\right)}$. The next step is the reinforcement
$\mathbf{E}_{n}$, which is effective with probability $c$, i.e.
$P\left(\mathbf{Z}_{n+1,i}=y|\mathbf{S}_{n}=s_{i},\mathbf{E}_{n}=y,\mathbf{Z}_{n,i}=z_{i}^{\left(n\right)}\right)=c$
and $P\left(\mathbf{Z}_{n+1,i}=z_{i}^{\left(n\right)}|\mathbf{S}_{n}=s_{i},\mathbf{E}_{n}=y,\mathbf{Z}_{n,i}=z_{i}^{\left(n\right)}\right)=1-c$.
The trial ends with a new (with probability $c$) state of conditioning
$\mathbf{Z}_{n+1}$.

\section{Oscillator model}

In this section we will describe intuitively the oscillator model.
We start by arguing for the use of neural oscillators as a way to
model the brain at a system level. We then discuss how we can represent
in a mathematically sensible way these oscillators. Finally, we show
how response computations and learning can be modeled using this theoretical
apparatus. Readers interested in more detail are referred to Suppes,
de Barros, \& Oas (2012). 

There are many different ways in which researchers try to figure out
how the brain works. For example, in cognitive neuroscience, among
the most popular research techniques are fMRI (functional magnetic
resonance imaging), MEG (magnetoencephalogram), and EEG (electroencephalogram).
MEG and EEG measure the electrical activities in the brain, whereas
fMRI measures changes in blood flow associated with higher metabolic
rates. While fMRI's popularity is due to its better spatial resolution,
MEG and EEG present significantly better time resolution. However,
what these techniques have in common is that, in order to measure
a signal from the brain, they require a large numbers of neurons to
fire synchronously. To make our point, let us focus on EEG (though
MEG would be adequate too). There are many experiments (see \citet{carvalhaes_using_2012}
and references) showing that the EEG data allow a good representation
of language or visual imagery. Thus, neurophysiological evidence points
toward language being an activity involving large collections of synchronizing
neurons, and we will center our model exactly on this. 

\begin{figure}
\begin{centering}
\includegraphics[scale=0.6]{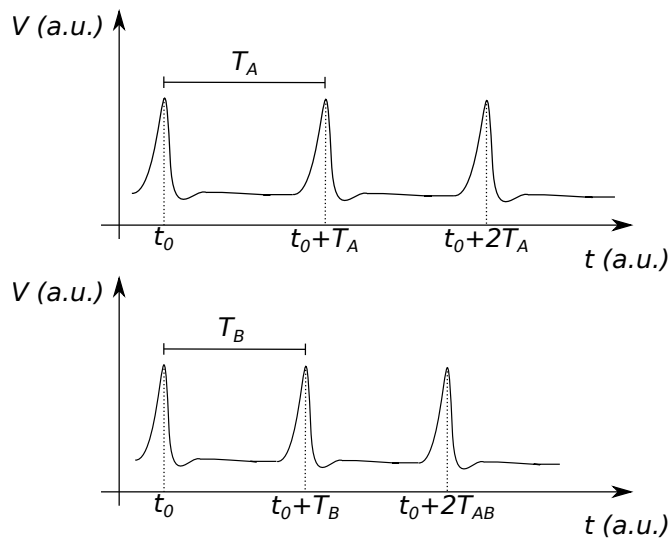}
\par\end{centering}

\caption{\label{fig:Approximate-action-potential}Approximate shape of the
action potentials $V_{A}$ and $V_{B}$ as a function of time $t$
for two uncoupled neurons $n_{A}$ and $n_{B}$ firing periodically,
with periods $T_{A}$ and $T_{B}$. For simplicity, we chose a $t_{0}$
when both neurons fire simultaneously. }

\end{figure}

Before we show how to describe such collections of synchronizing neurons
mathematically, it is useful to think about the physical mechanisms
of synchronization. Let us look first at individual neurons, and then
think about ensembles of neurons. Figure \ref{fig:Approximate-action-potential}
shows the qualitative behavior of two neurons $n_{A}$ and $n_{B}$
firing periodically, with $T_{B}<T_{A}$. 
 What happens if we now couple $n_{A}$ to an excitatory synapse coming
from neuron $n_{B}$? Because $t_{0}+T_{B}<t_{0}+T_{A}$, the excitatory
coupling will increase the membrane potential of neuron $n_{A}$ before
$t_{0}+T_{A}$, causing $n_{A}$ to fire a little earlier than it
would if it were not connected to $n_{B}$. So, excitatory synaptic
couplings between neurons can change the timing of coupling, and this
timing is changed such that the firings of both neurons approach (in
this case, the firing of $n_{A}$ approaches that of $n_{B}$). In
other words, excitatory couplings push $n_{A}$ and $n_{B}$ toward
synchronization. In fact, it is possible to prove mathematically that
if the number of neurons is large enough, the sum of the many weak
synaptic interactions can cause a strong effect, making all neurons
fire closer together \citep{izhikevich_dynamical_2007}; even when
weakly coupled, ensembles of periodically firing neurons synchronize.
It is interesting to note that the argument shown above can be scaled
up to distinct collections of neurons. Imagine we have two ensembles
of neurons, $N_{A}$ and $N_{B}$, such that neurons in them synchronize.
If neurons in $N_{A}$ and $N_{B}$ become coupled, then the same
mechanism as discussed above will be at play, and the ensembles will
synchronize among themselves. We will come back to this point later,
when we talk about response mechanisms. 

We are now in good shape to introduce the intuition behind the mathematical
description for the dynamics of synchronization. One of the main simplifying
assumptions we make is that the relevant information coded in the
brain is represented by the synchronization of an ensemble of neurons.
This ensemble may include tens of thousands of neurons, but because
they are synchronized, we can represent them, at least in first approximation,
by a single dynamical variable. To understand this, let us think about
the simplest case, where an oscillator $O\left(t\right)$ can be represented
by a sine function%
\footnote{We use a sine function for simplicity, but the following argument
is valid for periodic functions. %
}: 
\begin{equation}
O\left(t\right)=A\sin\omega t,\label{eq:simple-oscillator}
\end{equation}
where $\omega=\omega\left(t\right)$ is its time-dependent frequency.
Since $\omega$ may be a function of time, the value of $O\left(t\right)$
is completely determined by the argument of the sine, i.e. by $\varphi=\omega t$.
The quantity $\varphi$ is the \emph{phase }of the oscillator $O\left(t\right)=A\sin\varphi\left(t\right)$.
Since collections of firing neurons have very little variability in
its intensity (except, as we see below, when they interfere), we can
describe a neural oscillator by its phase. The interaction of a neural
oscillator with other neural oscillators may change the evolution
of its phase. 

We emphasize that there is a certain invariance of scale in the above
argument: it somehow does not matter how many neurons we have; all
that matters is that their amplitude does not vary, that their couplings
are strong enough to produce synchronization, and that their dynamics
is encoded in the phase. Furthermore, in the same way that individual
oscillating neurons synchronize to each other, a collection of coherent
neurons can also synchronize to another collection of coherent neurons.
Since neurons firing coherently may be described approximately by
their phase, we can focus on the phase dynamics, instead of being
concerned about the full description of the very complex dynamical
system. 

Now, let us look a little more into the details of the mathematics
of two synchronizing oscillators. Let us start with two oscillators,
$O_{1}\left(t\right)$ and $O_{2}\left(t\right)$, described by their
phases $\varphi_{1}$ and $\varphi_{2}$. If the two oscillators are
uncoupled and their frequency $\omega$ is constant, then it is clear
from equation (\ref{eq:simple-oscillator}) that they should satisfy
the following set of differential equations,
\begin{eqnarray}
\frac{d\varphi_{1}}{dt} & = & \omega_{1},\label{eq:uncoupled-oscillator-1}\\
\frac{d\varphi_{2}}{dt} & = & \omega_{2},\label{eq:uncoupled-oscillator-2}
\end{eqnarray}
where $\omega_{i}$, $i=1,2$, are their natural frequencies. However,
if they are weakly coupled, such that their interaction does not affect
the overall form of the oscillations given by $O_{1}\left(t\right)$
and $O_{2}\left(t\right)$ but affects their phase, then equations
(\ref{eq:uncoupled-oscillator-1}) and (\ref{eq:uncoupled-oscillator-2})
need to be modified to include changes to the phase. Furthermore,
if the underlying interaction is such that it will make the phases
approach each other, such as in the case of synaptically coupled neurons,
then it is possible to show that, in first approximation, the modified
dynamical equations become
\begin{eqnarray}
\frac{d\varphi_{1}}{dt} & = & \omega_{1}-k_{12}\sin\left(\varphi_{1}-\varphi_{2}\right),\label{eq:coupled-oscillator-1}\\
\frac{d\varphi_{2}}{dt} & = & \omega_{2}-k_{21}\sin\left(\varphi_{2}-\varphi_{1}\right),\label{eq:coupled-oscillator-2}
\end{eqnarray}
where $k_{ij}$ are the phase coupling strengths. If we extend this
to allow for $N$ oscillators, equations (\ref{eq:coupled-oscillator-1})
and (\ref{eq:coupled-oscillator-2}) then become
\begin{equation}
\frac{d\varphi_{i}}{dt}=\omega_{i}-\sum_{j\neq i}k_{ij}\sin\left(\varphi_{i}-\varphi_{j}\right).\label{eq:kuramoto-equation}
\end{equation}
Equation (\ref{eq:kuramoto-equation}) is known as Kuramoto equation
\citep{kuramoto_chemical_1984}, and it is widely used to describe
complex systems with emergent synchronization. The strength and usefulness
of Kuramoto's equation comes from two main points. First, it can be
solved under certain symmetric conditions and in the limit of large
$N$, yielding significant insight into the nature of emerging synchronization.
Second, a set of weakly-coupled oscillating dynamical systems close
to a Andronov-Hopf bifurcation can be described, in first approximation,
by Kuramoto-like equations (see \citet{izhikevich_dynamical_2007}).
For our purpose, Kuramoto's equations are a good approximation for
the dynamics of coupled neural oscillators. 

So, we now turn into the discussion of how we can think of stimulus
and response as modeled by oscillators, and in particular by Kuramoto's
equations. The basic idea is simple. Once a distal stimulus is presented,
the perceptual system activates an ensemble of brain neurons, $N_{s}$,
associated with it. This system itself is described by Kuramoto's
equations, and, because it synchronizes, we use its average phase
to describe its mean dynamics. If this stimulus elicits a response,
the activation of the response neurons via synaptic couplings follows.
Responses, as stimuli, are also represented by synchronously firing
ensemble of neurons. The selection of a particular response happens
when the stimulus oscillator synchronizes in phase with it, and such
phase is determined by the relative couplings between stimulus and
response oscillators. Let us now look more into its detail. 

The simplest stimulus-response neural oscillator model requires three
oscillators, $O_{s}$, $O_{r_{1}}$, and $O_{r_{2}}$. $O_{s}$ is
the oscillator representing firing neurons corresponding to the sampling
of a stimulus, and $O_{r_{1}}$ and $O_{r_{2}}$ are the response
oscillators. Their phases are $\varphi_{s}$, $\varphi_{r_{1}}$,
and $\varphi_{r_{2}}$. Before we describe their dynamics, let us
go through the process of a response computation. Whenever $O_{s}$
is activated, and subsequently $O_{r_{1}}$ and $O_{r_{2}}$, then
the intensity of firings (i.e., the rate of firing, as the individual
neuron amplitudes are reasonably stable) in each response oscillator
is not only due to its firing, but also to the firings of $O_{s}$.
As we mentioned earlier, a collection of firing neurons may interfere,
and in this case, interference means stronger firing rates when in
phase, and weaker firing rates when off of phase. Let us analyze this
with a mathematically simple example of equal intensity harmonic oscillators,
given by
\begin{eqnarray}
O_{s}(t) & = & A\cos\left(\omega_{0}t\right)=A\cos\left(\varphi_{s}(t)\right),\label{eq:oscillation-s}\\
O_{r_{1}}(t) & = & A\cos\left(\omega_{0}t+\delta\phi_{1}\right)=A\cos\left(\varphi_{r_{1}}(t)\right),\label{eq:oscillation-1}\\
O_{r_{2}}(t) & = & A\cos\left(\omega_{0}t+\delta\phi_{2}\right)=A\cos\left(\varphi_{r_{2}}(t)\right).\label{eq:oscillation-2}
\end{eqnarray}
Equations (\ref{eq:oscillation-s})--(\ref{eq:oscillation-2}) represent
the case where the oscillators are already synchronized with the same
frequency $\omega_{0}$ but with relative but constant phase differences
$\delta\phi_{1}$ and $\delta\phi_{2}$. The mean intensity give us
a measure of the excitation carried by the oscillations, and for the
superposition of $O_{s}(t)$ and $O_{r_{1}}(t)$ it is given by 
\begin{eqnarray*}
I_{1} & = & \left\langle \left(O_{s}(t)+O_{r_{1}}(t)\right)^{2}\right\rangle _{t}\\
 & = & \left\langle O_{s}(t)^{2}\right\rangle _{t}+\left\langle O_{r_{1}}(t)^{2}\right\rangle _{t}+\left\langle 2O_{s}(t)O_{r_{1}}(t)\right\rangle _{t},
\end{eqnarray*}
where $\left\langle f\left(t\right)\right\rangle _{t_{0}}=\frac{1}{\Delta T}\int_{t_{0}}^{t_{0}+\Delta T}f\left(t\right)dt$
($\Delta T\gg1/\omega_{0}$) is the time average . A quick computation
yields 
\[
I_{1}=A^{2}\left(1+\cos\left(\delta\phi_{1}\right)\right),
\]
and, similarly for $I_{2}$, 
\[
I_{2}=A^{2}\left(1+\cos\left(\delta\phi_{2}\right)\right).
\]
Therefore, the intensity depends on the phase difference between the
response-computation oscillators and the stimulus oscillator. 

Now, the maximum intensity of $I_{1}$ and $I_{2}$ is $2A^{2}$,
whereas their minimum intensity is zero. If we think of $I_{1}$ and
$I_{2}$ as competing possible responses, the maximum difference between
them happens when one of their relative phases (with respect to the
stimulus oscillator) is zero while the other is $\pi$. It is standard
to use the contrast, defined by 
\begin{equation}
b=\frac{I_{1}-I_{2}}{I_{1}+I_{2}},\label{eq:visibility}
\end{equation}
as a measure of how different the intensities are. From its definition,
$b$ takes values between $-1$ and $1$. When $I_{1}$ and $I_{2}$
are as different as possible, $\left|b\right|=1$; if, on the other
hand, $I_{1}$ and $I_{2}$ are the same, $b=0$. 

The contrast provides us with a useful way to think about responses
that are between $r_{1}$ and $r_{2}$. To see this, let us impose
\begin{equation}
\delta\phi_{1}=\delta\phi_{2}+\pi\equiv\delta\phi,\label{eq:ideal-phase-diff}
\end{equation}
which results in 
\begin{equation}
I_{1}=A^{2}\left(1+\cos\left(\delta\phi\right)\right),\label{eq:phase-1}
\end{equation}
 and
\begin{equation}
I_{2}=A^{2}\left(1-\cos\left(\delta\phi\right)\right).\label{eq:phase-2}
\end{equation}
In this case, the single parameter $\delta\phi$ is sufficient to
determine the contrast, as 
\begin{eqnarray}
b & = & \cos\left(\delta\phi\right),\label{eq:angle-reinforcement-b}
\end{eqnarray}
 $0\leq\delta\varphi\leq\pi$. So, the phase difference $\delta\phi$
between stimulus and response oscillators codes a continuum of responses
between $-1$ and $1$ (more precisely, because $\delta\varphi$ is
a phase, the interval is in the unit circle $\mathbb{T}$, and not
in a compact interval in $\mathbb{R}$). For arbitrary intervals $(\zeta_{1},\zeta_{2})$,
all that is required is a re-scaling of $b$. 

To summarize the above arguments. When a stimulus and response oscillators
activate, they fire periodically,leading to their synchronization
with constant phase relation. This phase relation causes interference,
which in turn determines the relative strength of the intensities
for each response. Thus, responses are determined by the interference
of oscillators, which is itself affected by the neural oscillators'
couplings. 

We now examine in more detail the mathematics of the stimulus and
response model. Let us look at each step of (\ref{eq:SR-trial}).

\subsubsection{Sampling}

When a stimulus $s_{n}$ is sampled, a collection of neurons start
firing synchronously, corresponding to the activation of a neural
oscillator, $O_{s_{n}}$. Such activation leads to a spreading of
activation to oscillators coupled to the stimulus oscillator, including
the response $O_{r_{1}}$ and $O_{r_{2}}$. Since the selection and
activation of $O_{s_{n}}$ involves the perceptual system, we do not
attempt to model with neural oscillators this step, but simply assume
their activation in a way that is consistent with the stochastic process
represented in SR theory by the random variable $\mathbf{S}_{n}$.
Furthermore, though it would be important to develop a detailed theory
of spreading activation, we do not, as for our current purposes it
suffices to simply assume the activation of $O_{r_{1}}$ and $O_{r_{2}}$.

\subsubsection{Response}

After the stimulus $s_{n}$ is sampled, the active oscillators evolve
for the time interval $\Delta t_{r}$, the time it takes to compute
a response, according to the following set of Kuramoto differential
equations. 
\begin{equation}
\frac{d\varphi_{i}}{dt}=\omega_{i}-\sum_{i\neq j}k_{ij}\sin\left(\varphi_{i}-\varphi_{j}+\delta_{ij}\right),\label{eq:Kuramoto-phase-differences}
\end{equation}
where $k_{ij}$ is the coupling constant between oscillators $i$
and $j$, and $\delta_{ij}$ is an anti-symmetric matrix representing
phase differences, and $i$ and $j$ can be either $O_{s_{n}}$, $O_{r_{1}}$,
or $O_{r_{2}}$. Here we use the notation where $O_{i}$ corresponds
to a neural oscillator and $\varphi_{i}$ to its phase. Equation (\ref{eq:Kuramoto-phase-differences})
can be rewritten as\textbf{ 
\begin{equation}
\frac{d\varphi_{i}}{dt}=\omega_{i}-\sum_{j}\left[k_{ij}^{E}\sin\left(\varphi_{i}-\varphi_{j}\right)+k_{ij}^{I}\cos\left(\varphi_{i}-\varphi_{j}\right)\right],\label{eq:kuramoto-equations-inhibition-excitation}
\end{equation}
}where $k_{ij}^{E}=k_{ij}\cos\left(\delta_{ij}\right)$ and $k_{ij}^{I}=k_{ij}\sin\left(\delta_{ij}\right)$,
which has an immediate physical interpretation: $k_{ij}^{E}$ corresponds
to excitatory couplings, whereas $k_{ij}^{I}$ corresponds to inhibitory
ones. These are the $4N$ excitatory and inhibitory coupling strengths
between oscillators. 
\begin{eqnarray}
\frac{d\varphi_{i}}{dt} & = & \omega_{0}-\sum_{i\neq j}\left[k_{i,j}^{E}\sin\left(\varphi_{i}-\varphi_{j}\right)-k_{i,j}^{I}\cos\left(\varphi_{i}-\varphi_{j}\right)\right],\label{eq:Kuramoto-inhib-assym}
\end{eqnarray}
where $\omega_{0}$ is their natural frequency. The solutions to (\ref{eq:Kuramoto-inhib-assym})
and the initial conditions randomly distributed at activation give
us the phases at time $t_{r,n}=t_{s,n}+\Delta t_{r}$. The coupling
strengths between oscillators determine their relative phase locking,
which in turn corresponds to the computation of a given response,
according to equation (\ref{eq:visibility}).

\subsubsection{Reinforcement and Conditioning}

As we saw above, the computation of a response depends on the inhibitory
and excitatory couplings between neural oscillators. Therefore, when
an effective reinforcement $\mathbf{Y}_{n}$ corresponding to changes
in the conditioning $\mathbf{Z}_{n+1}$ occurs, the coupling strengths
change. As with stimulus and responses, we represent a reinforcement
by a neural oscillator. Such oscillator, with frequency $\omega_{e}$,
is activated during reinforcement, and we assume that it forces the
reinforced response-computation and stimulus oscillators to synchronize
with the same phase difference of $\delta\varphi$, while the two
response-computation oscillators are kept synchronized with a phase
difference of $\pi$. Let the reinforcement oscillator be activated
on trial $n$ at time $t_{e,n}$, $t_{r,n+1}>t_{e,n}>t_{r,n}$, for
an interval of time $\Delta t_{e}$. Let $K_{0}$ be the coupling
strength between the reinforcement oscillator and the stimulus and
response-computation oscillators. In order to match the probabilistic
SR axiom governing the effectiveness of reinforcement, we also assume
that there is a normal probability distribution governing the coupling
strength $K_{0}$ between the reinforcement and the other active oscillators
with probability density 
\begin{equation}
f\left(K_{0}\right)=\frac{1}{\sigma_{K_{0}}\sqrt{2\pi}}\exp\left\{ -\frac{1}{2\sigma_{K_{0}}^{2}}\left(K_{0}-\overline{K}_{0}\right)^{2}\right\} .\label{eq:K0-density}
\end{equation}
When a reinforcement is effective, all active oscillators phase-reset
at $t_{e,n}$, and during reinforcement the phases of the active oscillators
evolve according to the following set of differential equations.
\begin{eqnarray}
\frac{d\varphi_{i}}{dt} & = & \omega_{0}-\sum_{i\neq j}\left[k_{i,j}^{E}\sin\left(\varphi_{i}-\varphi_{j}\right)-k_{i,j}^{I}\cos\left(\varphi_{i}-\varphi_{j}\right)\right]\nonumber \\
 &  & -K_{0}\sin\left(\varphi_{i}-\omega_{e}t+\Phi_{i}\right),\label{eq:learningphaseS-inhib-excite-first}
\end{eqnarray}
where $\Phi_{s_{n}}-\Phi_{r_{1}}=\delta\varphi$ and $\Phi_{r_{1}}-\Phi_{r_{2}}=\pi$.
The excitatory couplings are reinforced if the oscillators are in
phase with each other, according to the following equations. 
\begin{eqnarray}
\frac{dk_{i,j}^{E}}{dt} & = & \epsilon\left(K_{0}\right)\left[\alpha\cos\left(\varphi_{i}-\varphi_{j}\right)-k_{i,j}^{E}\right].\label{eq:learning-asym-excitatory-1}
\end{eqnarray}
Similarly, for inhibitory connections, if two oscillators are perfectly
off sync, then we have a reinforcement of the inhibitory connections.
\begin{eqnarray}
\frac{dk_{i,j}^{I}}{dt} & = & \epsilon\left(K_{0}\right)\left[\alpha\sin\left(\varphi_{i}-\varphi_{j}\right)-k_{i,j}^{I}\right],\label{eq:learning-asym-inhibitory-last}
\end{eqnarray}
In the above equations, 
\begin{equation}
\epsilon\left(K_{0}\right)=\left\{ \begin{array}{c}
0\mbox{ if }K_{0}<K'\\
\epsilon_{0}\mbox{ otherwise},
\end{array}\right.\label{eq:epsilon}
\end{equation}
where $\epsilon_{0}\ll\omega_{0}$, $\alpha$ and $K_{0}$ are constant
during $\Delta t_{e}$, and $K'$ is a threshold constant throughout
all trials. We can think of $K'$ as a threshold below which the reinforcement
oscillator has no (or very little) effect on the stimulus and response-computation
oscillators. For large enough values of $\Delta t_{e}$, the behavioral
probability parameter $c$ of effective reinforcement mentioned above
is, from (\ref{eq:K0-density}) and (\ref{eq:epsilon}), reflected
in the equation:
\begin{equation}
c=\int_{K'}^{\infty}f\left(K_{0}\right)\, dK_{0}.\label{eq:prob-eff-reinf}
\end{equation}
 This relationship comes from the fact that, if $K_{0}<K'$, there
is no effective learning from reinforcement, since there are no changes
to the couplings due to (\ref{eq:learning-asym-excitatory-1})--(\ref{eq:learning-asym-inhibitory-last}),
and (\ref{eq:Kuramoto-inhib-assym}) describing the oscillators' behavior.
Intuitively $K'$ is the effectiveness parameter: the larger it is,
the smaller the probability of effective reinforcement.

\section{Final remarks}

In this paper we described the neural oscillator model presented in
Suppes, de Barros, \& Oas (2012), with particular emphasis to the
physics and intuition behind many of the processes represented by
equations (\ref{eq:Kuramoto-inhib-assym}). To summarize it, the coded
phase differences were used to model a continuum of responses within
SR theory in the following way. At the beginning of a trial a stimulus
oscillator is activated, and with it the response oscillators. Then,
the coupled oscillator system evolves according to (\ref{eq:Kuramoto-inhib-assym})
if no reinforcement is present, and according to (\ref{eq:learningphaseS-inhib-excite-first})--(\ref{eq:learning-asym-inhibitory-last})
if reinforcement is present. The coupling constants and the conditioning
of stimuli are not reset at the beginning of each trial, and changes
to couplings correspond to effective reinforcement. Because of the
finite amount of time for a response, the probabilistic characteristics
of the initial conditions lead to the smearing of the phase differences
after a certain time, with an effect similar to that of the smearing
distribution in the SR model for a continuum of responses \citep{arrow_stimulus_1959}. 

We emphasize that in this paper we focused mainly on the physical
basis of our model, and did not go much into mathematical detail.
Furthermore, in Suppes, de Barros, \& Oas (2012) we applied the neural
oscillator model to many different experimental situations illustrated
in the literature, whereas here we did not address in detail any empirical
data. Interested readers are referred to our original paper. 

SR theory has enjoyed tremendous success in the past, and, in a certain
sense, its main features are still present in modern day neuroscience.
We believe that by showing how neurons may result in theoretical structures
that are somewhat similar to SR ones, as done in Suppes, de Barros,
\& Oas (2012), we can provide the basis for an extension of SR theory
that could be considered more realistic. For example, in our model,
many parameters, such as time of response, frequency of oscillations,
coupling strengths, etc., were fixed based on reasonable assumptions.
However, a more detailed and systematic study should be able to relate
such parameters to either underlying physiological constraints or
to behavioral variations, thus opening up the possibilities for new
empirical studies that go beyond SR theory. Also, in our model we
postulated many features without showing or proving their dynamics
from underlying neuronal dynamics. This was the case for the activation
of a stimulus and the spreading of activation of a stimulus and responses.
A more detailed theory based on neural oscillators of such dynamics
would certainly provide interesting empirical tests. 

Finally, the use of neural oscillators and interference may also help
explain certain aspects of cognition that are considered ``non-classical.''
The distinction between classical and quantum behavior is a subtle
one, and still not yet understood. For example, a well studied quantum-like
decision making process is the violation of Savage's sure-thing principle,
shown in a series of experiments by Tversky and Shaffir \citep{shafir_thinking_1992,tversky_disjunction_1992}.
Similar violations do not need any quantum-like representation in
the form of a Hilbert space, as proposed in the literature, but instead
can be obtained by interference of neural oscillators \citep{de_barros_quantum-like_2012}.
Furthermore, the use of neural oscillator interference even leads
to predictions that are not compatible with a Hilbert space structure
\citep{de_barros_joint_2012}, suggesting that the use of quantum-like
processes is not as quantum as many would wish. 

\bibliographystyle{apalike}
\bibliography{Suppes90thFetschrift}

\end{document}